\documentstyle[aps,preprint]{revtex}
\tightenlines

\begin{document}
\title{Calculation of Neel temperature for $S=1/2$ Heisenberg quasi-one-dimensional
antiferromagnets}
\author{V.Yu.Irkhin$^{*}$ and\ A.A.Katanin}
\address{620219, Institute of Metal Physics, Ekaterinburg, Russia.}
\maketitle

\begin{abstract}
Isotropic $S=1/2$ quasi-one-dimensional antiferromagnets are considered
within the bosonization method.\ The $1/z_{\perp }$-corrections to the
interchain mean-field theory (where $z_{\perp }$ is the number of nearest
neighbors in transverse to chain directions) are obtained for the
ground-state sublattice magnetization $\overline{S}_0$ and Neel temperature $%
T_N$. The corrections to $T_N$ make up about $25\%$ of mean-field value,
while those to $\overline{S}_0$ are small enough (especially in the
three-dimensional case). The fluctuation corrections obtained improve
considerably the agreement with the experimental data for magnetic-chain
compounds KCuF$_3,$ Sr$_2$CuO$_3$ and Ca$_2$CuO$_3.$
\end{abstract}


\section{Introduction}

Systems containing chains of magnetic atoms are investigated for a long time
from both theoretical and experimental point of view. There exist many real
compounds which are ``almost'' one-dimensional (1D), i.e. have small
interchain coupling. Here belong, e.g., KCuF$_3,$ Sr$_2$CuO$_3$ (spin $S=1/2$%
), CsNiCl$_3$ ($S=1$), CsVCl$_3$ ($S=3/2$) etc. There are a number of
approaches which give a possibility to perform calculations for purely 1D
magnets (Bethe ansatz, exact numerical diagonalization, different versions
of numerical renormalization group, quantum Monte-Carlo method etc.). At the
same time, consideration of multichain problem with the use of these methods
meets difficulties, so that theoretical approaches are of interest which can
adequately describe the situation in quasi-1D magnets in the presence of
interlayer coupling and/or anisotropy.

As for purely 1D antiferromagnets, there is well-known theoretical result by
Haldane \cite{Haldane} who mapped the spin-chain problem to nonlinear-sigma
model (NL$\sigma $M) and showed that the cases of integer and half-integer
spins differ qualitatively (for a review see, e.g., Ref. \cite{AffleckRev}).
For half-integer spins, the so-called topological $\theta $-term in the
effective action occurs which leads to unusual magnetic behavior of such
chains. As follows from the Bethe ansatz solution for $S=1/2$ (the same
situation holds for any half-integer spin value), ground-state in this case
already possesses quasi-long-range order. The excitation spectrum turns out
to be gapless and spin correlators have a power-law behavior, but staggered
magnetization is zero (the situation is reminiscent of the $XY$ model below
the Kosterlitz-Thouless point $T_{KT}$). It is natural to suppose that in
such a state the true long-range order is induced by an arbitrarily small
interchain coupling $J^{\prime }$ and/or magnetic anisotropy. For the
isotropic Heisenberg model, this problem was investigated within different
theoretical methods. The interchain mean-field theory \cite
{Scalapino,Schulz,Tsvelik} predicts for the ground-state staggered
magnetization $\overline{S}_0$ and Neel temperature $T_N$ the results
\begin{equation}
\overline{S}_0\propto \sqrt{|J^{\prime }|/J},T_N\propto |J^{\prime }|
\label{cr}
\end{equation}
and therefore indeed yields occurrence of long-range order at arbitrarily
small $|J^{\prime }|.$ The behavior (\ref{cr}) contradicts to the standard
spin-wave theory which does not distinguish between integer and half-integer
spins and predicts in both the cases a finite critical value, $J_c^{\prime
}\sim Je^{-\pi S},$ so that at $|J^{\prime }|<J_c^{\prime }$ the quantity $%
\overline{S}_0$ vanishes and
\begin{equation}
\overline{S}_0\propto \ln |J^{\prime }/J_c^{\prime }|,T_N\propto \overline{S}%
_0\sqrt{|J^{\prime }|}  \label{sw}
\end{equation}
for $|J^{\prime }|\ $greater but not too close to $J_c^{\prime }.$

This contradiction was resolved within the renormalization-group (RG)
approach \cite{AffleckRev,AffleckRG1,AffleckRG2,Wang} which showed that for
inverse-length scales $\mu \gg J_c^{\prime }/J$ the standard two-dimensional
NL$\sigma $M scaling equations are applicable, and the spin-field scale
factor $Z_\mu $ indeed satisfies $Z_\mu ^{-1/2}\propto \ln \mu .$ At the
same time, for half-integer spins at $\mu \ll J_c^{\prime }/J$ one has $%
Z_\mu ^{-1/2}\propto \mu ^{1/2}.$ This means\cite{AffleckRG1,AffleckRG2}
that for both integer and half-integer spins and $|J^{\prime }|\gg
J_c^{\prime }$ we have the spin-wave behavior (\ref{sw}), while for
half-integer spins and $|J^{\prime }|\ll J_c^{\prime }$ Eq.(\ref{cr}) holds.
(We suppose here that for half-integer spins the renormalized coupling
constant satisfies $g_\mu <g_c$ where $g_c$ is the critical 3D coupling
constant. Apparently, this inequality holds in the absence of dimerization,
see Refs. \cite{AffleckRG1,AffleckRG2}.)

In the extremely quantum case $S=1/2$ we have $J_c^{\prime }\sim J,$ so that
$|J^{\prime }|\ll J_c^{\prime }$ in a broad region of $|J^{\prime }|.$
Therefore, one can conclude that interchain mean-field theory of Ref.\cite
{Scalapino,Schulz,Tsvelik} gives a qualitatively correct description of $%
S=1/2$ quasi-1D magnets. At the same time, this theory does not take into
account interchain fluctuations. In particular, the calculated value of the
Neel temperature is not sensitive to space dimensionality of the system,
although in the $d=1+1$ case (both the dimensions are supposed to be
spatial, but second one corresponds to the direction, transverse to the
chain) we should have $T_N=0;$ for the $d=1+2$ case the values of $T_N$ turn
out to be too high in comparison with experimental data.

To obtain the corrections to interchain mean-field theory, we use the $%
1/z_{\perp }$-expansion ($z_{\perp }$ is the number of nearest neighbors in
directions transverse to the chain). This approach is similar to the
expansion in $1/z$ (or inverse interaction radius $1/R$), which has been
used to improve the standard mean-field theory of Heisenberg magnets many
years ago in Refs. \cite{VLP,Izyumov}. This approach is also equivalent to
the spin-fluctuation approach in the theory of itinerant magnets by Moriya
\cite{Moriya}.

The plan of paper is as follows. In Sect.2 we consider the bosonization of
the system of interacting Heisenberg chains. In Sect.3 we calculate
fluctuation corrections to the interchain mean-field theory. In Sect.3 we
discuss the results and compare them with experimental data on magnetic
chain compounds. In Appendix A the perturbation theory in $J^{\prime }$ is
considered and the first-order $1/z_{\perp }$ correction to the mean-field
value of Neel temperature are calculated. In Appendix B we demonstrate how
the same results can be obtained more elegantly in spirit of the
spin-fluctuation approach by Moriya. Finally, in Appendix C fluctuation
corrections to the ground-state staggered magnetization are derived.

\section{The model and its bosonization}

We consider the $S=1/2$ isotropic Heisenberg model of quasi-1D
antiferromagnet
\begin{equation}
{\cal H}=J\sum_{n,i}{\bf S}_{n,i}{\bf S}_{n+1,i}+\frac 12J^{\prime
}\sum_{n,<ij>}{\bf S}_{n,i}{\bf S}_{n,j}  \label{HH}
\end{equation}
where $n$ numerates sites along the chains and $i,j$ are indices of the
chains, $J>0$ and $J^{\prime }$ are intra- and interchain exchange
parameters respectively. We consider only the case $|J^{\prime }|\ll J.$

Each chain can be ``bosonized'' with the use of the standard relations (see,
e.g., Ref. \cite{TsvelikBook})
\begin{equation}
{\bf S}_{n,i}={\bf J}_i(x)+(-1)^n{\bf n}_i(x)
\end{equation}
where
\begin{eqnarray}
J_i^z(x) &=&\frac \beta {2\pi }\partial _x\varphi _i(x)  \nonumber \\
J_i^{\pm }(x) &=&\frac \lambda \pi \exp \left[ \pm i\beta \theta
_i(x)\right] \cos \beta \varphi _i(x)
\end{eqnarray}
are the cyclic vector current components and
\begin{eqnarray}
n_i^z(x) &=&\frac \lambda \pi \cos \beta \varphi _i(x)  \nonumber \\
n_i^{\pm }(x) &=&\frac \lambda \pi \exp \left[ \pm i\beta \theta _i(x)\right]
\end{eqnarray}
are their ``staggered'' analogs. Here $\lambda \ $is the scale
renormalization constant, $\varphi _i(x)$ is the boson operator, $\beta =%
\sqrt{2\pi }.$

Then we obtain the bosonized Hamiltonian in the form\cite{TsvelikL}
\begin{eqnarray}
{\cal H} &=&\frac v2\sum_i\int dx\left[ \Pi _i^2+(\partial _x\varphi
_i)^2\right] +g_u\sum_i\int dx\cos 2\beta \varphi _i  \nonumber \\
&&\ \ -\frac{J^{\prime }\lambda ^2}{2\pi ^2}\sum_{i,\delta _{\perp }}\int
dx\left[ \cos (\beta \varphi _i)\cos (\beta \varphi _{i+\delta _{\perp
}})+\cos \beta (\theta _{i+\delta _{\perp }}-\theta _i)\right]  \label{Ham}
\end{eqnarray}
where $v=\pi J/2$, $\Pi _i$ is the momentum that is canonically cojugated to
$\varphi _i,$ and $\theta _i$ satisfies $\partial _x\theta _i=-\Pi _i.$ The
first line in (\ref{Ham}) corresponds to a system of separate chains and has
the form of a standard sine-Gordon Hamiltonian. First term in (\ref{Ham})
describes a free-boson system, and the second one arises because of Umklapp
scattering of original fermions (which arises after applying the
Jordan-Wigner transformation); this term is marginal and produces
logarithmic corrections to thermodynamic quantities \cite
{Schulz,AffleckGSZ,Barzykin1,Barzykin2}. Calculations (see Refs. \cite
{Schulz,AffleckGSZ}) give $g_u/(2\pi )\simeq 0.25.$ The second line of (\ref
{Ham})\ describes the interaction between the chains. Note that only
relevant terms are included in the second line since the marginal terms give
smaller contribution (see Ref. \cite{TsvelikL}).

\section{Mean-field approximation for bosonized Hamiltonian and $1/
\lowercase{z}_{\perp} $-corrections}

The simplest way of treating interchain exchange interactions is the
mean-field approximation \cite{Schulz}. Decoupling the interaction term
\begin{equation}
\cos (\beta \varphi _i)\cos (\beta \varphi _{i+\delta _{\perp }})\rightarrow
2\langle \cos (\beta \varphi _{i+\delta _{\perp }})\rangle \cos (\beta
\varphi _i)
\end{equation}
we obtain
\begin{eqnarray}
{\cal H}_{MF} &=&\frac v2\sum_i\int dx\left[ \Pi _i^2+(\partial _x\varphi
_i)^2\right] +g_u\sum_i\int dx\cos 2\beta \varphi _i  \nonumber \\
&&\ \ \ \ \ \ \ \ \ \ \ \ \ \ \ \ \ \ \ -\frac \lambda \pi h_{MF}\sum_i\int
dx\cos (\beta \varphi _i)  \label{HMF}
\end{eqnarray}
where
\begin{equation}
h_{MF}=z_{\perp }J^{\prime }\lambda \langle \cos (\beta \varphi _i)\rangle
/\pi ,
\end{equation}
$z_{\perp }$ is the number of nearest neighbors in the transverse (to chain)
directions ($z_{\perp }=4$ for simple cubic lattice). This approximation
gives a possibility to reduce the multi-chain problem to a single-chain one
in an effective staggered magnetic field. Introducing the function
\begin{equation}
B(h;T)=\frac \lambda \pi \langle \cos (\beta \varphi _i)\rangle _h  \label{B}
\end{equation}
which should be calculated in the presence of the last term in (\ref{HMF}),
we obtain the self-consistent equation for the sublattice magnetization $%
\overline{S}$ in the mean-field approximation in the form
\begin{equation}
\overline{S}_{MF}=B(z_{\perp }J^{\prime }\overline{S}_{MF};T)
\end{equation}
Despite the Hamiltonian ${\cal H}_{MF}$ (\ref{HMF}) has an one-chain form,
calculation of the function $B(h;T)$ (which is an analog of the Brillouin
function in the usual mean-field theory of Heisenberg magnets) at arbitrary $%
T$ is a very complicated task. Scaling arguments suggest $%
B(h;T)=h^{1/3}f(h^{2/3}/T)$ with some scaling function $f(x).$ For $g_u=0$
(in this case we have a standard sine-Gordon, or, equivalently, massive
Thirring model) $B(h;T)$ was calculated by Bethe ansatz in Ref.\cite
{Chung-Chang}. However, in two following cases the calculation can be
performed analytically: (i) $T=0$ where we have\cite{Schulz,Tsvelik}
\begin{equation}
B(h;0)\simeq 0.677(h/v)^{1/3}\left[ 1+(g_u/2\pi )\ln (v/\Delta )\right]
^{1/2}
\end{equation}
where
\[
\Delta \simeq 2.085v^{1/3}h^{2/3}
\]
and (ii) $h\rightarrow 0$ where
\begin{equation}
B(h,T)=h\chi _0(T).  \label{BhT}
\end{equation}
Here $\chi _0(T)$ is the staggered susceptibility of the system in the
absence of $h$ \cite{Schulz,Barzykin2},
\begin{equation}
\chi _0(T)=\frac{\widetilde{\chi }_0}TL\left( \frac{\Lambda J}T\right) ,\;%
\widetilde{\chi }_0=\frac{\Gamma ^2(1/4)}{4\Gamma ^2(3/4)}\simeq 2.1184,
\label{Hi0T}
\end{equation}
where we have picked out the factor $\widetilde{\chi }_0$ for the sake of
convenience and
\begin{equation}
L(\Lambda J/T)=C\left[ \ln \frac{\Lambda J}T+\frac 12\ln \ln \frac{\Lambda J}%
T+{\cal O}(1)\right] ^{1/2}  \label{L}
\end{equation}
is the spin-field renormalization factor that arises because of the presence
of the marginal operator; the single-chain numerical calculations \cite
{Sandvik} yield $C\simeq 0.15,$ $\Lambda \simeq 5.8$. Thus one can see that
the above-mentioned scaling function $f(x)$ satisfies $f(x)\sim x$ at $%
x\rightarrow 0$ and $f(\infty )=$ const. The result (\ref{BhT})\ gives a
possibility to calculate the value of $T_N$ in the mean-field theory since
for $T\rightarrow T_N$ we just have $h_{MF}\rightarrow 0.$ Thus we obtain
the equation \cite{Schulz}
\begin{equation}
T_N^{MF}=z_{\perp }J^{\prime }\widetilde{\chi }_0L(\Lambda J/T_N^{MF})
\label{TNMF}
\end{equation}
We have included in (\ref{L}) a double-logarithmic term which was not taken
into account in Ref. \cite{Schulz} and modifies somewhat numerical results
(see below). As discussed in the Introduction, the mean-field approximation (%
\ref{TNMF}) is not quite satisfactory to describe experimental data. In
particular, the values of Neel temperatures are considerably overestimated.

The reason of this is that the mean-field approximation does not take into
account the collective excitations which substantially contribute to the
thermodynamic properties. Such excitations can be considered within the
random-phase approximation (RPA). The RPA spin susceptibilities are given by
\cite{Schulz,Tsvelik}
\begin{mathletters}
\label{HiRPA}
\begin{eqnarray}
\chi ^{+-}(q_z,\omega ) &=&\frac{\chi _0^{+-}(q_z,\omega )}{1-J^{\prime
}(q_x,q_y)\chi _0^{+-}(q_z,\omega )} \\
\chi ^{zz}(q_z,\omega ) &=&\frac{\chi _0^{zz}(q_z,\omega )}{1-J^{\prime
}(q_x,q_y)\chi _0^{zz}(q_z,\omega )}
\end{eqnarray}
where, for the square lattice in the direction transverse to chains,
\end{mathletters}
\begin{equation}
J^{\prime }(q_x,q_y)=2J^{\prime }(\cos q_x+\cos q_y),
\end{equation}
$\chi _0(q,\omega )$ being the dynamical staggered susceptibility for the
model (\ref{HMF}) and we have taken into account only staggered components
of the susceptibility. Again, $\chi _0(q,\omega )$ is given by simple
analytical expressions only in two cases: $T=0$ (for the results see Ref.%
\cite{Tsvelik} and also Appendix C), and $h\rightarrow 0$ where we have \cite
{SchulzB,Barzykin2} for both the susceptibilities
\begin{eqnarray}
\chi _0(q_z,\omega ) &=&\frac 1TL\left( \frac \Lambda T\right) \widetilde{%
\chi }_0(q_z/T,\omega /T)  \nonumber \\
\widetilde{\chi }_0(k,\nu ) &=&\frac 14\frac{\Gamma (1/4+ik_{+})\Gamma
(1/4+ik_{-})}{\Gamma (3/4+ik_{+})\Gamma (3/4+ik_{-})},\;k_{\pm }=\frac{\nu
\pm k}{4\pi }
\end{eqnarray}
As it should be, $\chi _0(0,0)=\chi _0(T).$

Now we can calculate the spin-fluctuation corrections to interchain
mean-field theory owing to collective modes. Similar to the case of the
simplest mean-field approximation in the theory of Heisenberg magnets \cite
{VLP,Izyumov}, these corrections can be obtained within the $1/z$-expansion.
Since we treat only transverse neighbors within the mean-field approach, one
has to speak about the $1/z_{\perp }$-expansion. To construct this
expansion, we consider the perturbation theory in $J^{\prime }/\max
(h_{MF},T)\sim 1/z_{\perp }$ (see Appendix A), which is an analog of
expansion in $J/\max (h_{MF},T)\sim 1/z$ for three-dimensional Heisenberg
magnets \cite{Izyumov}. From this viewpoint, the above-discussed mean-field
approximation is just the zeroth-order in $1/z_{\perp },$ so that the
fluctuation corrections to this approximation can be obtained in a regular
way. The leading (first order in $1/z_{\perp }$) corrections come from the
diagrams which include one RPA-interaction line.

The details of calculations are discussed in Appendix A. For the Neel
temperature we obtain to first order in $1/z_{\perp }$
\begin{equation}
T_N=kJ^{\prime }z_{\perp }\widetilde{\chi }_0L(\Lambda /T_N)  \label{TN}
\end{equation}
where
\begin{eqnarray}
k &=&\left\{ 1-\frac{\pi ^2}{2\widetilde{\chi }_0}\int\limits_{-\infty
}^\infty dr\int\limits_0^1d\tau \widetilde{V}(r,\tau )\left[ \frac 18%
F(r,\tau )+\frac 12G(r,\tau )\right] \right\} ^{-1}  \nonumber \\
\widetilde{V}(r,\tau ) &=&\int\limits_{-\infty }^\infty \frac{dq_z}{2\pi }%
\sum_n\sum_{q_x,q_y}\frac{\cos q_x+\cos q_y}{2\widetilde{\chi }_0-(\cos
q_x+\cos q_y)\widetilde{\chi }_0(q_z,2\pi in)}\exp (iq_zr-2\pi in\tau )
\label{V}
\end{eqnarray}
and $F(r,\tau )$, $G(r,\tau )$ are the four-point averages determined in
Appendix A. The result (\ref{TN}) differs from the mean-field result (\ref
{TNMF})\ by a factor of $k$ which depends only on the lattice structure in
the directions perpendicular to chains. Numerical calculation for $d=1+2$
case (simple cubic lattice) yields $k\simeq 0.70.$ Thus, with account of the
function $L(\Lambda /T_N),$ the lowering of $T_N$ due to interchain
fluctuation effects is about $25\%$ of its mean-field value. For $d=1+2$ the
integral in (\ref{V}) is divergent and we have $T_N=0.$

The same result (\ref{TN}) can also be derived in a more elegant way within
the spin-fluctuation approach by Moriya \cite{Moriya} (see Appendix B).

Corrections to the ground-state staggered magnetization are calculated in
Appendix C. We have
\begin{equation}
\overline{S}_0=(0.677-0.311I)h_{MF}^{1/3}  \label{S0}
\end{equation}
where the last term in the brackets represents the $1/z_{\perp }$ correction
with
\begin{equation}
I=\left\{
\begin{array}{cc}
0.038 & d=1+2 \\
0.193 & d=1+1
\end{array}
\right. .
\end{equation}
Note that we do no take into account the logarithmic corrections owing to
presence of the marginal operator, since there exists no simple ways of
calculating dynamical staggered susceptibility at $T=0$ in the presence of
such an operator. However, one can see that we have nearly $10\%$ lowering
of $\overline{S}_0$ for $d=1+1$ and only $2\%$ lowering for $d=1+2.$ Thus
the fluctuation corrections to ground-state magnetization are much less
important than those to the Neel temperature, and in the three-dimensional
case they can be neglected.

\section{Comparison with the experimental data and conclusion}

The results obtained enable one to perform quantitative comparison with
experimental data on magnetic chain systems. Consider first the compound KCuF%
$_3$ with $S=1/2$. According to Ref.\cite{KCuF3}, we have $J=406$ K, $%
\overline{S}_0/S=0.25.$ As discussed by Schulz \cite{Schulz}, this value of $%
\overline{S}_0$ corresponds to $J^{\prime }/J=0.047,$ so that $J^{\prime
}=19.1$ K. The simplest mean field approximation (\ref{TNMF}) yields $T_N=47$%
K. From (\ref{TN}) we obtain $T_N=37.7$ K which is somewhat lower in
comparison with the experimental result of Ref.\cite{KCuF3}, $T_N=39$ K.
Thus our approximation slightly overestimates the effects of fluctuations,
but improves reasonably the mean-field approximation. Contribution of the
double-logarithmic term in (\ref{L}) makes up about $5$ percents and
improves the agreement with the experimental data.

Another $S=1/2$ chain compound that is widely discussed in recent
publications is Sr$_2$CuO$_3$ which has the following parameters \cite
{SrCuO3,SrCa}: $J=2600$K, $T_N=5$K. Direct experimental data for $J^{\prime
} $ are absent, but using (\ref{TN})\ and the experimental value of $T_N$ we
obtain $J^{\prime }=1.85$K. Then we have from (\ref{S0}) $\overline{S}%
_0/S=0.042$ which is in agreement with the experimental data ($\overline{S}%
_0/S\leq 0.05$).

For Ca$_2$CuO$_3$ the experimental parameters have following values \cite
{SrCuO3,SrCa}: $S=1/2,$ $J=2600$K and $T_N=11$K. From (\ref{TN})\ we find $%
J^{\prime }=4.3$ K and $\overline{S}_0/S=0.062$. Taking into account above
results for Sr$_2$CuO$_3$ we find that the latter value is again in
excellent agreement with the experimental data \cite{SrCa} which give $%
\overline{S}_0$(Ca$_2$CuO$_3$)/$\overline{S}_0$(Sr$_2$CuO$_3$) $=1.5\pm 0.1$%
. Thus the result (\ref{TN}) is sufficient to describe quantitatively real
quasi-1D magnetic systems.

In the isotropic quasi-1D magnets under consideration, the fluctuation
corrections modify only numerical factor in the expression for $T_N.$ One
can expect, however, that in the anisotropic case the form of functional
dependence $T_N(J^{\prime })$ will be also modified. The influence of
anisotropy on the Neel temperature will be considered elsewhere. Another
interesting question concerns quasi-1D magnets with half-integer spins $%
S>1/2.$ As discussed in the Introduction, in this case there is a crossover
from ``usual'' spin-wave behavior of staggered magnetization to
non-spin-wave one. The expressions for $T_N$ should be also changed because
of this crossover.

Finally, despite the standard spin-wave theory yields a qualitatively
correct description of integer-spin magnetic chains, the corresponding
values of Neel temperatures are also overestimated in comparison with
experimental data. Thus calculation of fluctuation corrections for these
magnets is also of interest.

\section{Acknowledgements}

We are grateful to L.D.Landau Institute of Theoretical Physics for
hospitality at the conference in Chernogolovka (June, 1999) and providing
scientific literature on low-dimensional models.

\section{Appendix A. Perturbation theory in $J^{\prime }$ and the diagram
technique for spin operators}

In this Appendix we consider perturbation theory in $J^{\prime }$ for the
field theory with the Lagrangian
\begin{eqnarray}
{\cal L} &=&\frac v2\sum_i\int d^2{\rm x}(\partial \varphi
_i)^2+g_u\sum_i\int d^2{\rm x}\cos 2\beta \varphi _i-\frac \lambda \pi
h\sum_i\int d^2{\rm x}\cos (\beta \varphi _i)  \nonumber \\
&&\ \ \ \ \ \ \ -\frac{J^{\prime }\lambda ^2}{2\pi ^2}\sum_{i,\delta _{\perp
}}\int d^2{\rm x}\left[ \cos (\beta \varphi _i)\cos (\beta \varphi
_{i+\delta _{\perp }})+\cos \beta (\theta _{i+\delta _{\perp }}-\theta
_i)\right]   \label{LL}
\end{eqnarray}
which corresponds to the Hamiltonian (\ref{Ham}),\ the external staggered
magnetic field $h$ being introduced. In (\ref{Ham}) we have used the complex
coordinate ${\rm x}=x+iv\tau .$ Farther in this Appendix we use the system
of units where $v=1.$ Consider the calculation of staggered magnetization
\[
\overline{S}=\lambda \langle \cos (\beta \varphi _i)\rangle /\pi
\]
The perturbation theory in $J^{\prime }$ is constructed in a standard way
(see, e.g., Ref. \cite{Zinn-Justin}). To obtain the series in $J^{\prime }$
we write down the expression in the path integral formalism
\begin{equation}
\overline{S}=\frac \lambda \pi \frac{\int D\varphi \cos (\beta \varphi
_i(0))\exp (-{\cal L}[\varphi ])}{\int D\varphi \exp (-{\cal L}[\varphi ])}
\label{SPath}
\end{equation}
To zeroth order in $J^{\prime }$ (i.e. at $J^{\prime }=0$) we have ${\cal L}=%
{\cal L}_0$ and
\begin{equation}
\overline{S}_0=B(h;T)
\end{equation}
where the function $B$ was introduced in (\ref{B}). Expanding (\ref{SPath})
in $J^{\prime }$ we obtain
\begin{eqnarray}
\overline{S} &=&\frac \lambda \pi \frac{\int D\varphi \cos (\beta \varphi
_i(0))\exp (-{\cal L}_0[\varphi ])(1-{\cal L}_{int}+{\cal L}_{int}^2/2+...)}{%
\int D\varphi \exp (-{\cal L}_0[\varphi ])(1-{\cal L}_{int}+{\cal L}%
_{int}^2/2+...)}  \nonumber \\
\  &=&\frac \lambda \pi \langle \cos (\beta \varphi _i(0))(1-{\cal L}_{int}+%
{\cal L}_{int}^2/2+...)\rangle _{0,\text{conn}}  \label{SExp}
\end{eqnarray}
where we have denoted $\langle ...\rangle _0=\int D\varphi ....\exp (-{\cal L%
}_0[\varphi ])/\int D\varphi \exp (-{\cal L}_0[\varphi ])$ and
\begin{equation}
\langle \cos (\beta \varphi _i(0)){\cal L}_{int}^n\rangle _{0,\text{conn}%
}=\langle \cos (\beta \varphi _i(0)){\cal L}_{int}^n\rangle
_0-\sum_{m=0}^{n-1}\frac{(n!)^2}{m!(n-m)!}\langle \cos (\beta \varphi _i(0))%
{\cal L}_{int}^m\rangle _0\langle {\cal L}_{int}^{n-m}\rangle _0
\end{equation}
Each term in (\ref{SExp}) can be represented by its own diagram; the diagram
technique is the same as that for spin operators \cite{VLP,Izyumov} (some
elements of diagramm technique are shown on Fig.1). All diagrams are
classified by powers of the parameter $J^{\prime }/\max (h_{MF},T)\sim
1/z_{\perp }.$ Diagrams of Fig. 2 have zeroth order in $1/z_{\perp }$. The
summation of these diagrams leads to a shift of the external magnetic field
by the mean field:
\begin{equation}
h\rightarrow \widetilde{h}=h+h_{MF},\;h_{MF}=z_{\perp }J^{\prime }\overline{S%
}  \label{MFM}
\end{equation}
(The same result could be obtained by eliminating the mean-field term
directly in \ref{LL}). The diagrams of first order in $1/z_{\perp }$ (see
Fig. 3a) have one RPA-interaction line (Fig. 3b). These are directly
connected to the RPA susceptibilities (\ref{HiRPA}) by
\[
V^{+-,zz}({\bf q},\omega )=J^{\prime }(q_x,q_y)+[J^{\prime }(q_x,q_y)]^2\chi
^{+-,zz}(q_z,\omega )
\]
Thus we obtain
\begin{equation}
V^{+-,zz}({\bf q},\omega )\ =\frac{J^{\prime }(q_x,q_y)}{1+\delta -J^{\prime
}(q_x,q_y)\chi _0^{+-,zz}(q_z,\omega )}  \label{Vq}
\end{equation}
where
\begin{eqnarray}
\chi _0^{zz}(q_z,\omega ) &=&\frac{\lambda ^2}{\pi ^2}\int d^2{\rm x\,}%
\langle \cos \beta \varphi _i(0)\cos \beta \varphi _i({\rm x})\rangle
_{0,ir}\exp (-iq_zx+i\omega _n\tau ),  \nonumber \\
\chi _0^{+-}(q_z,\omega ) &=&\frac{\lambda ^2}{\pi ^2}\int d^2{\rm x}\langle
e^{i\beta [\theta _i(0)-\theta _i({\rm x})]}\rangle _0\exp (-iq_zx+i\omega
_n\tau ),  \label{Hi0q}
\end{eqnarray}
the two-operator irreducible average being given by
\begin{equation}
\langle AB\rangle _{ir}=\langle AB\rangle -\langle A\rangle \langle B\rangle
\label{Ir1}
\end{equation}
and, following to \cite{Moriya}, we have introduced the correction $\delta $
in the denominator to satisfy the self-consistency requirement. In the case $%
T\leq T_N$ under consideration, this is determined by the condition $[\chi
^{+-}(0,0)]^{-1}=0$, i.e. $\delta =z_{\perp }J^{\prime }\chi _0^{+-}(0,0)-1$%
. Transforming (\ref{Vq}) back to the real space,
\begin{equation}
V^{+-,zz}({\rm x})=T\sum_{i\omega _n}\int\limits_{-\pi }^\pi \frac{dq_z}{%
2\pi }\sum_{q_x,q_y}V^{+-,zz}({\bf q},i\omega _n)\exp (iq_zx-i\omega _n\tau )
\end{equation}
we obtain for the sublattice magnetization (see the diagramms of Fig.3a)
\begin{eqnarray}
\overline{S} &=&B(\widetilde{h};T)+\frac{\lambda ^3}{2\pi ^3}\int d^2{\rm x\,%
}d^2{\rm y}\left[ V^{zz}({\rm x}-{\rm y})\langle \cos \beta \varphi
_i(0)\cos \beta \varphi _i({\rm x})\cos \beta \varphi _i({\rm y})\rangle
_{0,ir}\right.   \nonumber \\
&&\ \ \ \ \ \ \ \ \ \ \ \ \ \ \ \ \ \ \ \ \left. V^{+-}({\rm x}-{\rm y}%
)\langle \cos \beta \varphi _i(0)e^{i\beta \theta _i({\rm x})}e^{-i\beta
\theta _i({\rm y})}\rangle _{0,ir}\right]   \label{MgFl}
\end{eqnarray}
where
\begin{equation}
\langle ABC\rangle _{ir}=\langle ABC\rangle -\langle A\rangle \langle
BC\rangle _{ir}-\langle B\rangle \langle AC\rangle _{ir}-\langle AB\rangle
_{ir}\langle C\rangle   \label{Ir2}
\end{equation}
(all averages are calculated with $h\rightarrow \widetilde{h}$). Up to this
moment, we did not use a concrete form of ${\cal L}_0.$ As already pointed
in the main text, the only case where the averages in (\ref{MgFl}) can be
calculated analytically is the limit $\widetilde{h}\rightarrow 0.$ In this
limit we have
\begin{eqnarray}
&&\ \ \ \ \ \ \langle \cos \beta \varphi _i(0)\cos \beta \varphi _i({\rm x}%
)\cos \beta \varphi _i({\rm y})\rangle _{0,ir}  \nonumber \\
\  &=&\frac \lambda \pi \widetilde{h}\int d^2{\rm z}\langle \cos \beta
\varphi _i(0)\cos \beta \varphi _i({\rm z})\cos \beta \varphi _i({\rm x}%
)\cos \beta \varphi _i({\rm y})\rangle _{0,ir}  \label{vh}
\end{eqnarray}
where
\begin{equation}
\langle ABCD\rangle _{ir}=\langle ABCD\rangle -\langle AD\rangle
_{ir}\langle BC\rangle _{ir}-\langle BD\rangle _{ir}\langle AC\rangle
_{ir}-\langle AB\rangle _{ir}\langle CD\rangle _{ir}  \label{Ir3}
\end{equation}
and similar expression for transverse components; the averages in the
right-hand side of (\ref{vh}) are calculated at $\widetilde{h}=0.$ Thus we
have at $h=0,$ $h_{MF}\rightarrow 0$%
\begin{eqnarray}
\overline{S} &=&\frac{\lambda ^2}{\pi ^2}h_{MF}\int d^2{\rm z}\langle \cos
\beta \varphi _i(0)\cos \beta \varphi _i({\rm z})\rangle   \nonumber \\
&&\ \ \ \ \ \ \ +\frac{\lambda ^4}{2\pi ^4}h_{MF}\int d^2{\rm x}\,d^2{\rm y}%
\,d^2{\rm z}\left[ V^{zz}({\rm x}-{\rm y})\langle \cos \beta \varphi
_i(0)\cos \beta \varphi _i({\rm x})\cos \beta \varphi _i({\rm y})\cos \beta
\varphi _i({\rm z})\rangle _{0,ir}\right.   \nonumber \\
&&\ \ \ \ \ \ \ \left. +V^{+-}({\rm x}-{\rm y})\langle \cos \beta \varphi
_i(0)\cos \beta \varphi _i({\rm z})e^{i\beta \theta _i({\rm x})-i\beta
\theta _i({\rm y})}\rangle _{0,ir}\right]   \label{SAv}
\end{eqnarray}
Note that the SU(2) invariance guarantees at $\widetilde{h}=0$%
\begin{eqnarray}
&&\ \ \int d^2{\rm x}\,d^2{\rm y}\,d^2{\rm z}\left[ \langle \cos \beta
\varphi _i(0)\cos \beta \varphi _i({\rm x})\cos \beta \varphi _i({\rm y}%
)\cos \beta \varphi _i({\rm z})\rangle _{0,ir}\right.   \nonumber \\
&&\ \ \left. -3\langle \cos \beta \varphi _i(0)\cos \beta \varphi _i({\rm x}%
)e^{i\beta \theta _i({\rm y})-i\beta \theta _i({\rm z})}\rangle
_{0,ir}\right]
\begin{array}{c}
=
\end{array}
0
\end{eqnarray}
Calculating at $\beta ^2=2\pi $ the averages in the right-hand side of (\ref
{SAv}) in the presence of the marginal operator $g_u\cos 2\beta \varphi _i$
(which produces logarithmic corrections) we obtain
\begin{eqnarray}
\overline{S} &=&\frac 12h_{MF}L\left( \frac \Lambda T\right) \int d^2{\rm z}%
\frac 1{|\varsigma ({\rm z})|}+\frac 1{16}h_{MF}L^2\left( \frac \Lambda T%
\right)   \nonumber \\
&&\ \ \times \int d^2{\rm x}\,d^2{\rm y}\,d^2{\rm z}V^{zz}({\rm x}-{\rm y}%
)\left[ \frac{|\varsigma ({\rm z})\varsigma ({\rm x}-{\rm y})|}{|\varsigma (%
{\rm x})\varsigma ({\rm y})\varsigma ({\rm z}-{\rm x})\varsigma ({\rm z}-%
{\rm y})|}+\frac{|\varsigma ({\rm z})\varsigma ({\rm z}-{\rm y})|}{%
|\varsigma ({\rm z})\varsigma ({\rm y})\varsigma ({\rm z}-{\rm x})\varsigma (%
{\rm x}-{\rm y})|}\right.   \nonumber \\
&&\ \ \left. +\frac{|\varsigma ({\rm y})\varsigma ({\rm z}-{\rm x})|}{%
|\varsigma ({\rm z})\varsigma ({\rm x})\varsigma ({\rm z}-{\rm y})\varsigma (%
{\rm x}-{\rm y})|}-\frac 2{|\varsigma ({\rm z})\varsigma ({\rm x}-{\rm y})|}-%
\frac 2{|\varsigma ({\rm x})\varsigma ({\rm y}-{\rm z})|}-\frac 2{|\varsigma
({\rm y})\varsigma ({\rm x}-{\rm z})|}\right]   \nonumber \\
&&\ \ +\frac 14h_{MF}L^2(\frac \Lambda T)\int d^2{\rm x}\,d^2{\rm y}\,d^2%
{\rm z}V^{+-}({\rm x}-{\rm y})\frac 1{|\varsigma ({\rm z})\varsigma ({\rm x}-%
{\rm y})|}  \nonumber \\
&&\ \times \text{Re}\left[ \sqrt{\frac{\varsigma ({\rm x})\varsigma ({\rm z}-%
{\rm y})\varsigma (\overline{{\rm z}}-\overline{{\rm x}})\varsigma (%
\overline{{\rm y}})}{\varsigma (\overline{{\rm x}})\varsigma (\overline{{\rm %
z}}-\overline{{\rm y}})\varsigma ({\rm z}-{\rm x})\varsigma ({\rm y})}}%
-1\right]
\end{eqnarray}
where
\begin{equation}
\varsigma ({\rm x})=\sinh (\pi T{\rm x})/(\pi T)
\end{equation}
and $L(\Lambda /T)$ is determined by (\ref{L}). Introducing ${\rm r}={\rm x}-%
{\rm y}$ instead of ${\rm x}$ and passing to the variables $\widetilde{{\rm r%
}}={\rm r}T$ etc. we obtain the result
\begin{eqnarray}
\overline{S}=\frac 1Th_{MF}\widetilde{\chi }_0L\left( \frac \Lambda T\right)
\left\{ 1+\frac{\pi ^2}{2T\widetilde{\chi }_0}L\left( \frac \Lambda T\right)
\int d^2{\rm r}V({\rm r})\left[ \frac 18F({\rm r})+\frac 12G({\rm r})\right]
\right\}   \label{MgF}
\end{eqnarray}
where
\begin{equation}
\widetilde{\chi }_0=\frac \pi 2\int d^2{\rm z}\frac 1{|\widetilde{\varsigma }%
({\rm z})|}\simeq 2.1184  \label{H0}
\end{equation}
and
\begin{eqnarray}
F({\rm r}) &=&\int d^2{\rm y}\,d^2{\rm z}\left[ \frac{|\widetilde{\varsigma }%
({\rm z})\widetilde{\varsigma }({\rm r})|}{|\widetilde{\varsigma }({\rm r}+%
{\rm y})\widetilde{\varsigma }({\rm y})\widetilde{\varsigma }({\rm z}-{\rm %
y-r})\widetilde{\varsigma }({\rm z}-{\rm y})|}+\frac{|\widetilde{\varsigma }(%
{\rm z})\widetilde{\varsigma }({\rm z}-{\rm y})|}{|\widetilde{\varsigma }(%
{\rm z})\widetilde{\varsigma }({\rm y})\widetilde{\varsigma }({\rm z}-{\rm %
y-r})\widetilde{\varsigma }({\rm r})|}\right.   \nonumber \\
&&\ \left. +\frac{|\widetilde{\varsigma }({\rm y})\widetilde{\varsigma }(%
{\rm z}-{\rm y}-{\rm r})|}{|\widetilde{\varsigma }({\rm z})\widetilde{%
\varsigma }({\rm r+y})\widetilde{\varsigma }({\rm z}-{\rm y})\widetilde{%
\varsigma }({\rm r})|}-\frac 2{|\widetilde{\varsigma }({\rm z})\widetilde{%
\varsigma }({\rm r})|}-\frac 2{|\widetilde{\varsigma }({\rm r}+{\rm y})%
\widetilde{\varsigma }({\rm y}-{\rm z})|}-\frac 2{|\widetilde{\varsigma }(%
{\rm y})\widetilde{\varsigma }({\rm r}+{\rm y}-{\rm z})|}\right]   \label{F}
\\
G({\rm r}) &=&\int \,d^2{\rm y}\,d^2{\rm z}\frac 1{|\widetilde{\varsigma }(%
{\rm z})\widetilde{\varsigma }({\rm r})|}\text{Re}\left[ \sqrt{\frac{%
\widetilde{\varsigma }({\rm r}+{\rm y})\widetilde{\varsigma }({\rm z}-{\rm y}%
)\widetilde{\varsigma }(\overline{{\rm z}}-\overline{{\rm y}}-\overline{{\rm %
r}})\widetilde{\varsigma }(\overline{{\rm y}})}{\widetilde{\varsigma }(%
\overline{{\rm r}}+\overline{{\rm y}})\widetilde{\varsigma }(\overline{{\rm z%
}}-\overline{{\rm y}})\widetilde{\varsigma }({\rm z}-{\rm y-r})\widetilde{%
\varsigma }({\rm y})}}-1\right]   \label{G}
\end{eqnarray}
In (\ref{H0})-(\ref{G}) we have used
\begin{equation}
\widetilde{\varsigma }({\rm x})=\sinh (\pi {\rm x})
\end{equation}
Finally, using the connection (\ref{MFM})\ between the mean field and the
staggered magnetization and collecting all corrections to the denominator
analogously to the usual three-dimensional Heisenberg magnets\cite{HM}, we
obtain the result\ (\ref{TN}) of the main text.

\section{Appendix B. Spin-fluctuation approach to thermodynamics of quasi-1D
Heisenberg magnets.}

The results of previous Appendix can be obtained in a much more simple way
with the use of the spin-fluctuation approach proposed by T. Moriya \cite
{Moriya} for description of thermodynamics of itinerant magnets where the
Stoner theory (which is an analog of the mean-field theory in Heisenberg
magnets) turns out to be quite not satisfactory. To apply the
spin-fluctuation approach, we represent Hamiltonian (\ref{HH}) as
\begin{eqnarray}
{\cal H} &=&{\cal H}_0+{\cal H}_{int} \\
{\cal H}_0 &=&J\sum_{n,i}{\bf S}_{n,i}{\bf S}_{n+1,i}-h_{MF}%
\sum_{n,i}(-1)^{n+i}S_{i,n}^z  \nonumber \\
{\cal H}_{int} &=&\frac 12J^{\prime }\sum_{<ij>}({\bf S}_{n,i}{\bf S}%
_{n,j})_{ir}  \nonumber
\end{eqnarray}
where $(AB)_{ir}=AB-\langle A\rangle \langle B\rangle .$ With the use of the
Hellmann-Feynman theorem, we obtain for the free energy
\begin{equation}
{\cal F}={\cal F}_0(h_{MF})+\frac 12\sum_{{\bf q},i\omega _n}\int dJ^{\prime
}\left[ \chi ^{+-}(q,i\omega _n)+\chi ^{zz}(q,i\omega _n)\right]
\end{equation}
where ${\cal F}_0(h_{MF})$ is the free energy corresponding to ${\cal H}_0.$
Using the RPA results (\ref{HiRPA})\ one can find
\begin{equation}
{\cal F}={\cal F}_0(h_{MF})+\frac 12\sum_{{\bf q},i\omega _n}\left\{ \ln
\left[ 1+J^{\prime }(q_x,q_y)\chi _0^{+-}(q_z,i\omega _n)\right] +\ln \left[
1+J^{\prime }(q_x,q_y)\chi _0^{zz}(q_z,i\omega _n)\right] \right\}
\end{equation}
Differentiating with respect to $h_{MF}$ we readily obtain
\begin{eqnarray}
\overline{S} &=&\overline{S}_{MF}+\frac 12\sum_{{\bf q},i\omega _n}\left[
\frac{J^{\prime }(q_x,q_y)}{1+J^{\prime }(q_x,q_y)\chi _0^{+-}(q_z,i\omega
_n)}\frac{\partial \chi _0^{+-}(q_z,i\omega _n)}{\partial h}\right.
\nonumber \\
&&\ \ \ \left. +\frac{J^{\prime }(q_x,q_y)}{1+J^{\prime }(q_x,q_y)\chi
_0^{zz}(q_z,i\omega _n)}\frac{\partial \chi _0^{zz}(q_z,i\omega _n)}{%
\partial h}\right]  \label{SF}
\end{eqnarray}
This is just the result (\ref{MgFl}) of Appendix A. Representing $\chi
_0(q_z,i\omega _n)$ via boson variables, differentiating in $h$ and
calculating again the corresponding averages we return to (\ref{MgF}).

\section{Appendix C. Ground-state fluctuation corrections in the absence of
marginal operator.}

In this Appendix we consider ground-state corrections to the mean-field
value of sublattice magnetization. We use the expression (\ref{MgFl}), or,
equivalently, (\ref{SF}) where, at $T=0$ (Refs. \cite{Schulz,Tsvelik}), 
\begin{eqnarray}
\chi _0^{+-} &=&\frac 1{4|J^{\prime }|}\frac{\Delta ^2}{\omega
^2+v^2q^2+\Delta ^2} \\
\chi _0^{zz} &=&\frac{Z^{\prime }/Z}{4|J^{\prime }|}\frac{\Delta ^2}{\omega
^2+v^2q^2+3\Delta ^2}
\end{eqnarray}
with 
\begin{eqnarray}
\Delta &\simeq &6.175\,|J^{\prime }|,Z^{\prime }/Z\simeq 0.49,  \nonumber \\
\overline{S}_0 &\simeq &1.017|J^{\prime }|
\end{eqnarray}
and $h_{MF}=z_{\perp }J^{\prime }\overline{S}_0$ (we neglect here the
contribution of marginal operator). Differentiating (\ref{SF}) in $h_{MF}$
(with account of implicit dependence of $J^{\prime }$ on $h_{MF}$) we obtain
after some algebraic manipulations 
\begin{eqnarray}
\overline{S} &=&\overline{S}_0-\frac \Delta {4\pi }\frac{\partial \Delta }{%
\partial h}I  \nonumber \\
I &=&\sum_q\left[ (1-\Gamma _q^{\prime }/2)\ln \frac 1{1-\Gamma _q^{\prime }}%
+(3-Z^{\prime }\Gamma _q^{\prime }/2Z)\ln \frac 1{1-Z^{\prime }\Gamma
_q^{\prime }/(3Z)}\right]
\end{eqnarray}
where we have used $\Gamma _q^{\prime }=\cos q$ for $d=1+1$ and $\Gamma
_q^{\prime }=(\cos q_x+\cos q_y)/2$ for $d=1+2.$ Calculating the integral $I$
numerically we obtain the result (\ref{S0}) of the main text.

{\sc Figure captions.}

Fig.1. Some elements of diagram technique for spin operators (for a detailed
description see Ref. \cite{Izyumov}). The first three irreducible averages
are determined by (\ref{Ir1}), (\ref{Ir2}) and (\ref{Ir3}).

Fig.2. Diagramms for staggered magnetization to zeroth order in $1/z_{\perp
} $ (mean-field approximation).

Fig.3. (a) Diagrams of first order in $1/z_{\perp }$ for staggered
magnetization (b) Equations for RPA interaction lines.

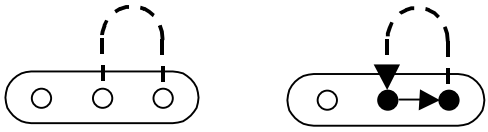
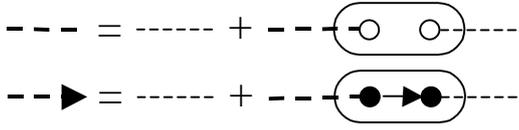
\end{document}